# Communication and Round Efficient Information Checking Protocol

Arpita Patra    C. Pandu Rangan

Dept of Computer Science and Engineering

Indian Institute of Technology, Madras

Chennai, India 600036

arpitapatra_10@yahoo.co.in, prangan55@yahoo.com

**Abstract**

In this paper, we present a very important primitive called Information Checking Protocol (ICP) which plays an important role in constructing statistical Verifiable Secret Sharing (VSS) and Weak Secret Sharing (WSS) protocols. Informally, ICP is a tool for authenticating messages in the presence of *computationally unbounded* corrupted parties. Here we extend the basic bare-bone definition of ICP, introduced by Rabin et al. [3] and then present an ICP that attains the best communication complexity and round complexity among all the existing ICPs in the literature. We also show that our ICP satisfies several interesting properties such as linearity property which is an important requirement in many applications of ICP.

Though not presented in this paper, we can design communication and round efficient statistical (i.e involves negligible error probability in computation) VSS and Multiparty Computation (MPC) protocol using our new ICP.

## 1 Introduction

### 1.1 Existing Literature and Existing Definition of ICP

The notion of ICP was first introduced by Rabin et al. [3]. Rabin et al. [3] have used ICP for constructing a statistical WSS protocol which was further used to design a statistical VSS protocol. Since then many ICPs have been designed [3, 1, 2] and used in constructing various statistical VSS [3, 1, 2] and WSS [3, 1, 2] protocols.

As described in [3, 1, 2], an ICP is executed among three parties: a *dealer D*, an *intermediary INT* and a *verifier R*. The dealer $D$ hands over a secret value $s$ to $INT$. At a later stage, $INT$ is required to hand over $s$ to $R$ and convince $R$ that $s$ is indeed the value which $INT$ received from $D$.

### 1.2 Our New Definition of ICP

The basic definition of ICP involves only a *single* verifier $R$ [3, 2, 1]. We extend this notion to *multiple* verifiers, specifically to $n$ verifiers/parties denoted by $\mathcal{P} = \{P_1, \ldots, P_n\}$ out of which at most $t$ are corrupted by unbounded powerful active adversary. Moreover $D$ and $INT$ are some specific party from $\mathcal{P}$. Thus our ICP is executed among three entities: a dealer $D \in \mathcal{P}$, an intermediary $INT \in \mathcal{P}$ and the entire set $\mathcal{P}$ acting as verifiers. Moreover, in contrast to the existing ICPs that deal with single secret, our ICP can deal with *multiple* secrets *concurrently* and thus achieves better communication complexity than multiple executions of ICP dealing with single secret.

The multiple secret, multiple receiver ICP is useful in the design of efficient protocols for statistical VSS and WSS. Statistical VSS is possible iff $n \geq 2t+1$ (provided a physical broadcast channel is available in the system) and for the design of statistical VSS with optimal resilience, we work with $n = 2t+1$. As our ICP is useful in such context, we design our ICP as well with $n = 2t+1$. Thus our ICP can be used for statistical VSS and WSS and they can be used for statistical MPC with optimal resilience (i.e $n = 2t+1$).

## 1.3 Our Network and Adversary Model

We consider a setting with $n$ parties (we also call them as verifiers) $\mathcal{P} = \{P_1, P_2, \ldots, P_n\}$ with $n = 2t+1$, that are pairwise connected by a secure (or private) channel. We further assume that all parties have access to a common broadcast channel (that allows any party in $\mathcal{P}$ to send some information identically to all other parties in $\mathcal{P}$). We assume the system to be synchronous. Therefore the protocols operate in a sequence of rounds, where in each round, a party performs some local computation, sends new messages to the other parties through the private channels and broadcasts some information over the broadcast channel, then it receives the messages that were sent by the other parties in this round on the private and broadcast channels.

The adversary that we consider is a *static, threshold, active and rushing* adversary having *unbounded computing power*. The adversary, denoted by $\mathcal{A}_t$, can corrupt at most $t$ parties out of the $n$ parties. The adversary controls and coordinates the actions of the corrupted/faulty parties in any arbitrary manner. We further allow the adversary to be *rushing* [2], i.e. in every round of communication it can wait to hear the messages of the honest parties before sending his own messages. We consider a static adversary, who corrupts all the parties at the beginning of the protocol.

We assume that the messages sent through the channels are from a specified domain. Thus if a party receives a message which is not from the specified domain (or a party receives no message at all), then he replaces it with some pre-defined default message. Thus, we separately do not consider the case when no message or syntactically incorrect message is received by a party.

## 1.4 Structure of ICP

As in [3, 1], our ICP is also structured into sequence of following three phases:

1. **Generation Phase**: This phase is initiated by $D$. Here $D$ hands over the secret $S$ containing $\ell$ elements from $\mathbb{F}$ (working field of ICP) to *intermediary* $INT$. In addition, $D$ sends some *authentication information* to $INT$ and *verification information* to individual verifiers in $\mathcal{P}$.

2. **Verification Phase**: This phase is initiated by $INT$ to acquire an IC Signature on $S$ that will be later accepted by every honest verifier in $\mathcal{P}$. Depending on the behavior of $D/INT$, secret $S$ OR $S$ along with the *authentication information*, held by $INT$ at the end of **Verification Phase** will be called as $D$'s IC *signature* on $S$ and will be denoted by $ICSig(D, INT, \mathcal{P}, S)$.

3. **Revelation Phase**: This phase is carried out by $INT$ and the verifiers in $\mathcal{P}$. Here $INT$ reveals $ICSig(D, INT, \mathcal{P}, S)$. The verifiers publish their responses after verifying $ICSig(D, INT, \mathcal{P}, S)$ with respect to their verification information. Depending upon the responses of the verifiers, every verifier $P_i \in \mathcal{P}$ either accepts $ICSig(D, INT, \mathcal{P}, S)$ or rejects it.

## 1.5 The properties of ICP

Our ICP satisfies the following properties (which are almost same as the properties, satisfied by the ICP of [3, 2]). In these properties, $\epsilon$ is called the error parameter.

1. **ICP-Correctness1:** If $D$ and $INT$ are *honest*, then $ICSig(D, INT, \mathcal{P}, S)$ will be accepted in **Revelation Phase** by each *honest* verifier.

2. **ICP-Correctness2:** If $INT$ is *honest* then at the end of **Verification Phase**, $INT$ possesses an $ICSig(D, INT, \mathcal{P}, S)$, which will be accepted in **Revelation Phase** by all honest verifiers, except with probability $\epsilon$.

3. **ICP-Correctness3:** If $D$ is *honest*, then during **Revelation Phase**, with probability at least $(1-\epsilon)$, every $ICSig(D, INT, \mathcal{P}, S')$ with $S' \neq S$, produced by a *corrupted* $INT$ will be rejected by *honest* verifiers.

4. **ICP-Secrecy:** If $D$ and $INT$ are *honest* then till the end of **Verification Phase**, $S$ is information theoretically secure from $\mathcal{A}_t$ (that controls $t$ verifiers in $\mathcal{P}$).

## 1.6 The Road-map

In section 2, we present our novel ICP with its complete proof. In section 3, we compare our ICP with the existing ICPs and show that our ICP attains the best communication and round complexity among all existing ICPs. Section 4 introduces a definition and a notation for our ICP. Section 5 then concentrates on the linearity property of our ICP. Finally, we conclude this article in section 6.

## 2 Our Novel ICP

In this section, we present an ICP called as MVMS-ICP (MVMS stands for Multi Verifier Multi Secret). Protocol MVMS-ICP requires one round for **Generation Phase** and two rounds for **Verification Phase** and **Revelation Phase** each.

To bound the error probability by $\epsilon$, our protocol MVMS-ICP operates over field $\mathbb{F} = GF(2^\kappa)$, where $\epsilon \geq n2^{-\kappa}$. Hence we have $|\mathbb{F}| \geq \frac{n}{\epsilon}$. Moreover we assume that $n = poly(\log \frac{1}{\epsilon})$. Now each element from the field is represented by $\kappa = \log |\mathbb{F}| = \mathcal{O}(\log \frac{n}{\epsilon}) = \mathcal{O}(\log n + \log \frac{1}{\epsilon}) = \mathcal{O}(\log \frac{1}{\epsilon})$ bits (the last equality in the above sequence follows from our assumption that $n = poly(\log \frac{1}{\epsilon})$). We now present an informal idea of MVMS-ICP.

**The Intuition:** In MVMS-ICP, $D$ selects a random polynomial $F(x)$ of degree $\ell + t$, whose lower order $\ell$ coefficients are the elements of $S$ and delivers $F(x)$ to $INT$. In addition, $D$ privately delivers to each individual verifier $P_i$, the value of $F(x)$ at a random, secret *evaluation point* $\alpha_i$. This distribution of information by $D$ helps to achieve **ICP-Correctness3** property. The reason is that if $D$ is *honest*, then a *corrupted* $INT$ cannot produce an *incorrect* $F'(x) \neq F(x)$ during **Revelation Phase** without being detected by an *honest* verifier with very high probability. This is because a corrupted $INT$ will have no information about the evaluation point of an honest verifier and hence with very high probability, $F'(x)$ will not match with $F(x)$ at the evaluation point held by an honest verifier.

The above distribution by $D$ also maintains **ICP-Secrecy** property. This is because the degree of $F(x)$ is $\ell + t$. But only up to $t$ points on $F(x)$ will be known to $\mathcal{A}_t$ through $t$ corrupted verifiers. Therefore $\mathcal{A}_t$ will fall short by $\ell$ points to *uniquely* interpolate $F(x)$.

But the above distribution alone is not enough to achieve **ICP-Correctness2**. A *corrupted* $D$ might distribute $F(x)$ to $INT$ and value of some other polynomial (different from $F(x)$) to each honest verifier. To detect this situation, $INT$ and the verifiers interact in *zero knowledge* fashion to check the consistency of $F(x)$ held by $INT$ and the values held by individual verifiers. The specific details of the zero knowledge, along with other formal steps of protocol MVMS-ICP are given in Fig. 1.

We now prove the properties of protocol MVMS-ICP.

**Claim 1** *If $D$ and $INT$ are honest then $D$ will never broadcast $S$ during Ver.*

PROOF: Since $INT$ is honest, he will correctly broadcast $(d, B(x))$ during **Round 1** of Ver. So during **Round 2** of Ver, $D$ will find $B(\alpha_i) = dv_i + r_i$ for all $i = 1, \ldots, n$. Thus $D$ will never broadcast $S$ during Ver. □

**Lemma 1 (ICP-Correctness1)** *If $D$ and $INT$ are honest, then $ICSig(D, INT, \mathcal{P}, S)$ produced by $INT$ during Reveal will be accepted by each honest verifier.*

PROOF: If $D$ is honest, then $(F(x), R(x))$ held by honest $INT$ and $(\alpha_i, v_i, r_i)$ held by honest verifier $P_i$ will satisfy $v_i = F(\alpha_i)$ and $r_i = R(\alpha_i)$. Moreover by Claim 1, $D$ will never broadcast $S$ during Ver. Hence $ICSig(D, INT, \mathcal{P}, S) = F(x)$. Now every honest verifier $P_i$ will broadcast `Accept` in **Round 2** of Reveal as condition **C1** i.e $v_i = F(\alpha_i)$ will hold. Since there are at least $t + 1$ honest verifiers, $ICSig(D, INT, \mathcal{P}, S)$ will be accepted by every honest verifier. □

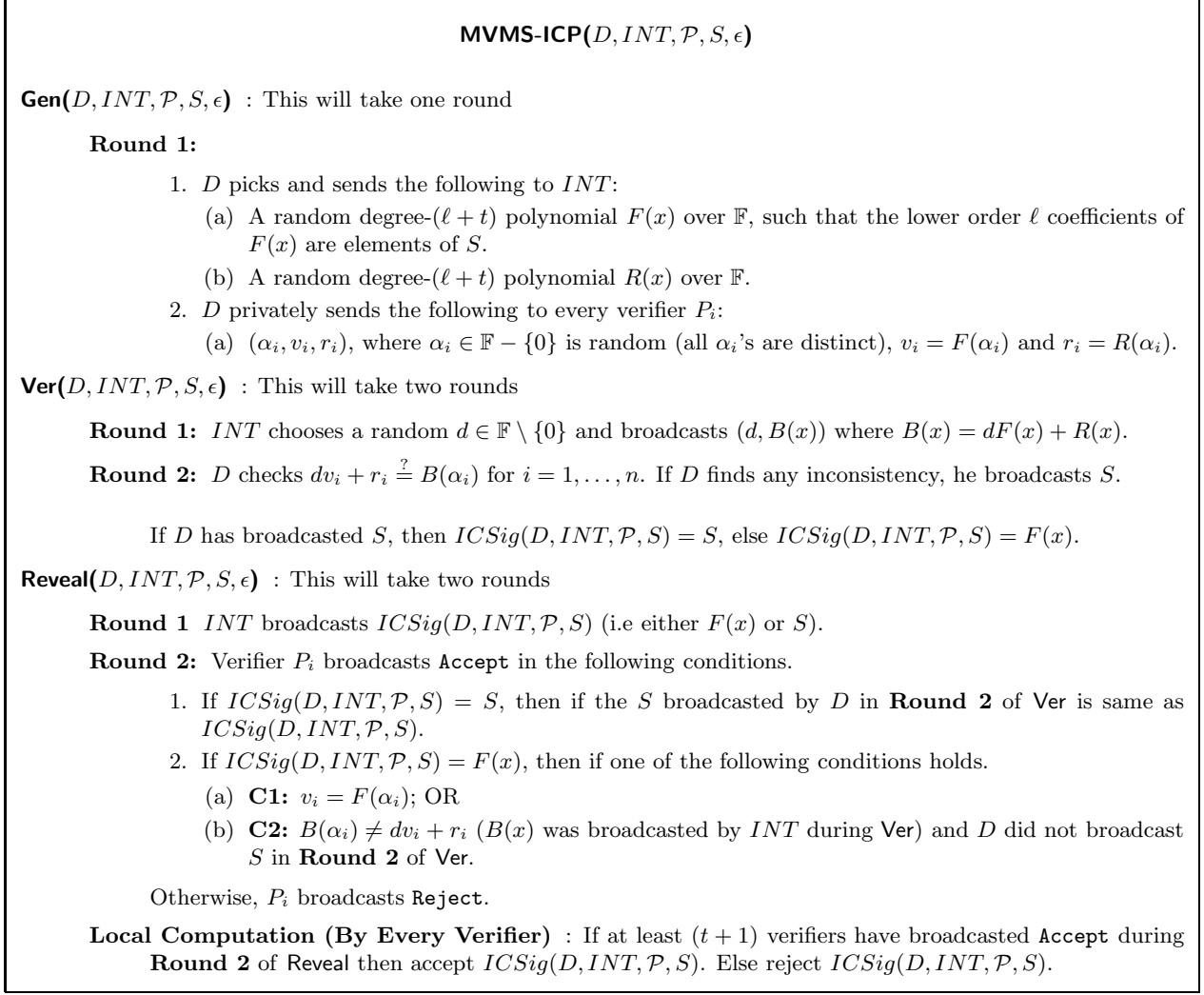

Figure 1: Protocol MVMS-ICP with $n = 2t + 1$ Verifiers

**MVMS-ICP$(D, INT, \mathcal{P}, S, \epsilon)$**

**Gen$(D, INT, \mathcal{P}, S, \epsilon)$** : This will take one round

**Round 1:**

1. $D$ picks and sends the following to $INT$:
   (a) A random degree-$(\ell + t)$ polynomial $F(x)$ over $\mathbb{F}$, such that the lower order $\ell$ coefficients of $F(x)$ are elements of $S$.
   (b) A random degree-$(\ell + t)$ polynomial $R(x)$ over $\mathbb{F}$.
2. $D$ privately sends the following to every verifier $P_i$:
   (a) $(\alpha_i, v_i, r_i)$, where $\alpha_i \in \mathbb{F} - \{0\}$ is random (all $\alpha_i$'s are distinct), $v_i = F(\alpha_i)$ and $r_i = R(\alpha_i)$.

**Ver$(D, INT, \mathcal{P}, S, \epsilon)$** : This will take two rounds

**Round 1:** $INT$ chooses a random $d \in \mathbb{F} \setminus \{0\}$ and broadcasts $(d, B(x))$ where $B(x) = dF(x) + R(x)$.

**Round 2:** $D$ checks $dv_i + r_i \stackrel{?}{=} B(\alpha_i)$ for $i = 1, \ldots, n$. If $D$ finds any inconsistency, he broadcasts $S$.

If $D$ has broadcasted $S$, then $ICSig(D, INT, \mathcal{P}, S) = S$, else $ICSig(D, INT, \mathcal{P}, S) = F(x)$.

**Reveal$(D, INT, \mathcal{P}, S, \epsilon)$** : This will take two rounds

**Round 1** $INT$ broadcasts $ICSig(D, INT, \mathcal{P}, S)$ (i.e either $F(x)$ or $S$).

**Round 2:** Verifier $P_i$ broadcasts Accept in the following conditions.

1. If $ICSig(D, INT, \mathcal{P}, S) = S$, then if the $S$ broadcasted by $D$ in **Round 2** of Ver is same as $ICSig(D, INT, \mathcal{P}, S)$.
2. If $ICSig(D, INT, \mathcal{P}, S) = F(x)$, then if one of the following conditions holds.
   (a) **C1:** $v_i = F(\alpha_i)$; OR
   (b) **C2:** $B(\alpha_i) \neq dv_i + r_i$ ($B(x)$ was broadcasted by $INT$ during Ver) and $D$ did not broadcast $S$ in **Round 2** of Ver.

Otherwise, $P_i$ broadcasts Reject.

**Local Computation (By Every Verifier)** : If at least $(t+1)$ verifiers have broadcasted Accept during **Round 2** of Reveal then accept $ICSig(D, INT, \mathcal{P}, S)$. Else reject $ICSig(D, INT, \mathcal{P}, S)$.

**Claim 2** *If $D$ is corrupted and $(F(x), R(x))$ held by an honest $INT$ and $(\alpha_i, v_i, r_i)$ held by an honest verifier $P_i$ satisfies $F(\alpha_i) \neq v_i$ and $R(\alpha_i) \neq r_i$, then except with probability $\frac{\epsilon}{n}$, $B(\alpha_i) \neq dv_i + r_i$.*

PROOF: We first prove that for $(F(x), R(x))$ held by an honest $INT$ and $(\alpha_i, v_i, r_i)$ held by honest verifier $P_i$, there is *only one* non-zero $d$ for which $B(\alpha_i) = dv_i + r_i$, even though $F(\alpha_i) \neq v_i$ and $R(\alpha_i) \neq r_i$. For otherwise, assume there exists another non-zero element $e \neq d$, for which $B(\alpha_i) = ev_i + r_i$ is true, even if $F(\alpha_i) \neq v_i$ and $R(\alpha_i) \neq r_i$. This implies that $(d-e)F(\alpha_i) = (d-e)v_i$ or $F(\alpha_i) = v_i$, which is a contradiction. Now since $d$ is randomly chosen by honest $INT$ only after $D$ handed over $(F(x), R(x))$ to $INT$ and $(\alpha_i, v_i, r_i)$ to $P_i$, a corrupted $D$ has to guess $d$ in advance during Gen to make sure that $B(\alpha_i) = dv_i + r_i$ holds. However, $D$ can guess $d$ with probability at most $\frac{1}{|\mathbb{F}|-1} \approx \frac{\epsilon}{n}$. Hence only with probability at most $\frac{\epsilon}{n}$, corrupted $D$ can ensure $B(\alpha_i) = dv_i + r_i$, even though $F(\alpha_i) \neq v_i$ and $R(\alpha_i) \neq r_i$. $\square$

**Lemma 2 (ICP-Correctness2)** *If $INT$ is honest then at the end of Ver, $INT$ possesses an $ICSig$ $(D, INT, \mathcal{P}, S)$, which will be accepted in Reveal by all honest verifiers, except with probability $\epsilon$.*

PROOF: We consider the case when $D$ is corrupted, because when $D$ is honest, the lemma follows from Lemma 1. Now the proof can be divided into following two cases:

1. $ICSig(D, INT, \mathcal{P}, S) = S$: This implies that $D$ has broadcasted $S$ during **Round 2** of Ver. In this case, the lemma holds trivially, without any error. This is because the honest $INT$ will correctly broadcast $ICSig(D, INT, \mathcal{P}, S) = S$ during **Round 1** of Reveal and every honest verifier will find that $S$ broadcasted by $INT$ is same as the one that was broadcasted by $D$ during **Round 2** of Ver. So all honest verifiers (at least $t+1$) will broadcast Accept and hence $ICSig(D, INT, \mathcal{P}, S)$ will be accepted by all honest verifiers.

2. $ICSig(D, INT, \mathcal{P}, S) = F(x)$: This implies that $D$ has not broadcasted anything during **Round 2** of Ver. Here, we first show that except with probability $\frac{\epsilon}{n}$, each honest verifier will broadcast Accept during Reveal. So let $P_i$ be an honest verifier. We have now the following cases depending on the relation that holds between the information held by $INT$ (i.e $(F(x), R(x))$) and information held by the honest $P_i$ (i.e $(\alpha_i, v_i, r_i)$):

    (a) If $F(\alpha_i) = v_i$: Here $P_i$ will broadcast Accept without any error probability as condition **C1** (i.e $F(\alpha_i) = v_i$) will hold.
    
    (b) If $F(\alpha_i) \neq v_i$ and $R(\alpha_i) = r_i$: Here $P_i$ will broadcast Accept without any error probability, as condition **C2** (i.e $B(\alpha_i) \neq dv_i + r_i$) will hold.
    
    (c) If $F(\alpha_i) \neq v_i$ and $R(\alpha_i) \neq r_i$: Here $P_i$ will broadcast Accept except with probability $\frac{\epsilon}{n}$, as condition **C2** will hold, except with probability $\frac{\epsilon}{n}$ (see Claim 2).

As shown above, there is a negligible error probability of $\frac{\epsilon}{n}$ with which an *honest* $P_i$ may broadcast Reject when $F(\alpha_i) \neq v_i$ and $R(\alpha_i) \neq r_i$ (i.e the third case). This happens if a corrupted $D$ can guess the unique $d$ in Gen, corresponding to $P_i$ and it so happens that $INT$ also selects the same $d$ in Ver and therefore condition **C2** does not hold good for $P_i$ in Reveal. Now $D$ can guess a $d_i$ for each honest verifier $P_i$ and if it so happens that honest $INT$ chooses $d$ which is same as one of those $t + 1$ $d_i$'s guessed by $D$, then condition **C2** will not be satisfied for the honest verifier $P_i$ for whom $d_i = d$ and therefore $P_i$ will broadcast Reject. This may lead to the rejection of $ICSig(D, INT, \mathcal{P}, S)$, as $t$ corrupted verifiers may always broadcast Reject. But the above event can happen with error probability $\frac{t+1}{|\mathbb{F}|-1} = (t+1)\frac{\epsilon}{n} \approx \epsilon$. This is because there are $t+1$ $d_i$'s and $INT$ has selected some $d$ randomly from $\mathbb{F} \setminus \{0\}$. This implies that all honest verifiers will broadcast Accept during Reveal, except with error probability $\epsilon$.

This completes the proof of the lemma. □

**Lemma 3 (ICP-Correctness3)** *If $D$ is honest then during Reveal, with probability at least $1 - \epsilon$, every $ICSig(D, INT, \mathcal{P}, S')$ with $S' \neq S$ revealed by a corrupted $INT$ will be rejected by honest verifiers.*

PROOF: Here again we have the following two cases:

1. $ICSig(D, INT, \mathcal{P}, S) = S$: This implies that $D$ has broadcasted $S$ during **Round 2** of Ver. In this case if a corrupted $INT$ tries to reveal $ICSig(D, INT, \mathcal{P}, S')$ where $S' \neq S$ then all honest verifiers (at least $t+1$) will broadcast Reject during Reveal. This is because the honest verifiers will find that $S'$ is not same as $S$ which was broadcasted by $D$ during **Round 2** of Ver.

2. $ICSig(D, INT, \mathcal{P}, S) = F(x)$: This implies that $D$ has not broadcasted anything during **Round 2** of Ver. Here a corrupted $INT$ can produce $S' \neq S$ by broadcasting $F'(x) \neq F(x)$ during Reveal such that the lower order $\ell$ coefficients of $F'(x)$ is $S'$. We now claim that if $INT$ does so, then except with probability $\frac{\epsilon}{n}$, an honest verifier $P_i$ will broadcast Reject during Reveal. In the following, we show that the conditions for which the honest verifier $P_i$ would broadcast Accept are either impossible or may happen with probability $\frac{\epsilon}{n}$:

    (a) $F'(\alpha_i) = v_i$: Since $P_i$ and $D$ are honest, corrupted $INT$ has no information about $\alpha_i, v_i$. Hence the probability that $INT$ can ensure $F'(\alpha_i) = v_i = F(\alpha_i)$ is same as the probability with which $INT$ can correctly guess $\alpha_i$, which is at most $\frac{1}{|\mathbb{F}-1|} \approx \frac{\epsilon}{n}$ (since $\alpha_i$ is randomly chosen by $D$ from $\mathbb{F}$).

(b) $B(\alpha_i) \neq dv_i + r_i$: This case is never possible because $D$ is honest. If $B(\alpha_i) \neq dv_i + r_i$ corresponding to $P_i$, then honest $D$ would have broadcasted $S$ during **Round 2** of Ver and hence $ICSig(D, INT, \mathcal{P}, S)$ would have been equal to $S$, which is a contradiction to our assumption that $ICSig(D, INT, \mathcal{P}, S) = F(x)$.

As shown above, there is a negligible error probability of $\frac{\epsilon}{n}$ with which an honest $P_i$ may broadcast Accept, even if the corrupted $INT$ produces $F'(x) \neq F(x)$. This happens if the corrupted $INT$ can guess $\alpha_i$ corresponding to honest verifier $P_i$. Now there are $t+1$ honest verifiers. A corrupted $INT$ can guess $\alpha_i$ for any one of those $t+1$ honest verifiers and thereby can ensure that $F'(\alpha_i) = v_i$ holds for some honest $P_i$ (which in turn implies $P_i$ will broadcast Accept). This will ensure that $INT$'s $ICSig(D, INT, \mathcal{P}, S')$ will be accepted, as $t$ corrupted verifiers may always broadcast Accept. But the above event can happen with probability at most $\frac{t+1}{|\mathbb{F}|-1} = (t+1)\frac{\epsilon}{n} \approx \epsilon$. This asserts that every $ICSig(D, INT, \mathcal{P}, S')$ with $S' \neq S$, revealed by a corrupted $INT$ will be rejected by all honest verifiers with probability at least $(1-\epsilon)$. □

**Lemma 4 (ICP-Secrecy)** *If $D$ and $INT$ are honest, then till the end of Ver, $S$ is information theoretically secure from $\mathcal{A}_t$ (that controls $t$ verifiers in $\mathcal{P}$).*

PROOF: During Gen, $\mathcal{A}_t$ will know $t$ distinct points on $F(x)$ and $R(x)$. Since both $F(x)$ and $R(x)$ are of degree-$(\ell + t)$, the lower order $\ell$ coefficients of both $F(x)$ and $R(x)$ are information theoretically secure. During Ver, $\mathcal{A}_t$ will know $d$ and $dF(x) + R(x)$. Since both $F(x)$ and $R(x)$ are random and independent of each other, the lower order $\ell$ coefficients of $F(x)$ remain to be information theoretically secure. Also, if $D$ and $INT$ are honest, then $D$ will never broadcast $S$ during Ver (from Claim 1). Hence the lemma. □

**Theorem 1** *Protocol MVMS-ICP is an efficient ICP.*

PROOF: Follows from Lemma 1, 2, 3 and 4. □

**Theorem 2 (Round Complexity of MVMS-ICP)** *In protocol MVMS-ICP, Gen requires one round, Ver and Reveal requires two rounds each.*

**Theorem 3 (Communication Complexity of MVMS-ICP)** *Protocol MVMS-ICP attains the following bounds: (a) Protocol Gen privately communicates $\mathcal{O}((\ell + n) \log \frac{1}{\epsilon})$ bits. (b) Protocol Ver and Reveal requires broadcast of $\mathcal{O}((\ell + n) \log \frac{1}{\epsilon})$ bits each.*

PROOF: In protocol Gen, $D$ privately gives $\ell + t$ field elements to $INT$ and three field elements to each verifier. Since each field element can be represented by $\kappa = \mathcal{O}(\log \frac{1}{\epsilon})$ bits, Gen incurs a private communication of $\mathcal{O}((\ell + n) \log \frac{1}{\epsilon})$ bits. In protocol Ver, $INT$ broadcasts $B(x)$ containing $\ell + t$ field elements, thus incurring broadcast of $\mathcal{O}((\ell + n) \log \frac{1}{\epsilon})$ bits. Moreover, $D$ may broadcast $S$ which will incur broadcast of $\mathcal{O}(\ell \log \frac{1}{\epsilon})$ bits. Therefore, in total Ver requires broadcast of $\mathcal{O}((\ell+n) \log \frac{1}{\epsilon})$ bits. In protocol Reveal, $INT$ broadcasts $F(x)$, consisting of $\ell+t$ field elements, while each verifier broadcasts `Accept/Reject` signal. So Reveal involves broadcast of $\mathcal{O}((\ell + n) \log \frac{1}{\epsilon})$ bits. □

## 3 Comparison of MVMS-ICP with the ICPs of [3] and [2]

Both the ICPs of [3] and [2] are designed in single verifier and single secret model. But they can be extended to the case of multiple (i.e. $n$) verifiers easily. Indeed in [3, 2], the single verifier ICPs were executed in parallel for $n$ verifiers in the implementation of VSS protocols. Moreover, as the protocols were designed for single secret, they can be extended for $\ell$ secrets by $\ell$ parallel invocations of the protocols. Since protocol MVMS-ICP is designed to handle $n$ verifiers and $\ell$ secrets concurrently, in Table 1, we compare our MVMS-ICP with the ICPs of [3] and [2] extended for $n$ verifiers and $\ell$ secrets.

Table 1: Communication Complexity and Round Complexity of protocol MVMS-ICP and Existing ICP with $n = 2t + 1$ verifiers and $\ell$ secrets.

| Ref. | Communication Complexity in Bits | | | Round Complexity | | |
|---|---|---|---|---|---|---|
| | Gen | Ver | Reveal | Gen | Ver | Reveal |
| [3] | Private– $\mathcal{O}(\ell n (\log \frac{1}{\epsilon})^2)$ | Broadcast– $\mathcal{O}(\ell n (\log \frac{1}{\epsilon})^2)$ | Broadcast– $\mathcal{O}(\ell n (\log \frac{1}{\epsilon})^2)$ | 1 | at least 3 | 2 |
| [2] | Private– $\mathcal{O}(\ell n \log \frac{1}{\epsilon})$ | Broadcast– $\mathcal{O}(\ell n \log \frac{1}{\epsilon})$ | Broadcast– $\mathcal{O}(\ell n \log \frac{1}{\epsilon})$ | 1 | 3 | 2 |
| This paper MVMS-ICP | Private– $\mathcal{O}((\ell + n) \log \frac{1}{\epsilon})$ | Broadcast– $\mathcal{O}((\ell + n) \log \frac{1}{\epsilon})$ | Broadcast– $\mathcal{O}((\ell + n) \log \frac{1}{\epsilon})$ | 1 | 2 | 2 |

## 4 Few Remarks, Definitions and Notations on ICP

### 4.1 MVMS-ICP with One Round of Reveal

It is interesting to note that if we restrict the adversary $\mathcal{A}_t$ to a non-rushing adversary then the two rounds of Reveal can be collapsed into a single round where $INT$ broadcasts $ICSig(D, INT, \mathcal{P}, S)$ and simultaneously every verifiers broadcast their values $(\alpha_i, v_i, r_i)$. It is easy to check that all the properties of ICP will hold in such a case. But in the presence of rushing adversary, the two rounds are needed in order to force a corrupted $INT$ to commit to the polynomial $F(x)$ prior to seeing the evaluation points, as this knowledge can enable the adversary to publish a polynomial that can match with the values broadcasted by the honest verifiers, which would violate the **ICP-Correctness3** property of the protocol. However, if the adversary is non-rushing then this property is achieved via the synchronicity of the step. Hence, we have the following theorem:

**Theorem 4** *If the adversary is non-rushing then there exists an efficient ICP with one round in Gen, two rounds in Ver and one round in Reveal.*

### 4.2 A Definition and a Notation

**Definition 1 (IC Signature with $\epsilon$ Error)** *An IC signature $ICSig(D, INT, \mathcal{P}, S)$ for some secret $S$, is said to have $\epsilon$ error, if it satisfies the following: 1. **ICP-Correctness1** without any error; 2. **ICP-Correctness2** with error probability of at most $\epsilon$; 3. **ICP-Correctness3** with error probability of at most $\epsilon$; 4. **ICP-Secrecy** without any error.*

Notice that if an IC signature is generated in MVMS-ICP (which is executed with error parameter $\epsilon$), then the IC signature will have $\epsilon$ error. This follows from the proofs of Lemma 1, 2, 3 and 4.

**Notation 1 (Notation for Using MVMS-ICP)** *We say that: 1. "$D$ sends $ICSig(D, INT, \mathcal{P}, S)$ having $\epsilon$ error to $INT$" to mean that $D$ executes $\mathsf{Gen}(D, INT, \mathcal{P}, S, \epsilon)$; 2. "$INT$ receives $ICSig(D, INT, \mathcal{P}, S)$ having $\epsilon$ error from $D$" to mean that the parties have executed $\mathsf{Ver}(D, INT, \mathcal{P}, S, \epsilon)$; 3. "$INT$ reveals $ICSig(D, INT, \mathcal{P}, S)$ having $\epsilon$ error" to mean that $\mathsf{Reveal}(D, INT, \mathcal{P}, S, \epsilon)$ has been executed.*
*Clearly if $D$ sends $ICSig(D, INT, \mathcal{P}, S)$ to $INT$ in $i^{th}$ round, then $INT$ will receive $ICSig(D, INT, \mathcal{P}, S)$ in $(i + 2)^{th}$ round, as $\mathsf{Ver}$ requires two rounds.*

## 5 Linearity of Protocol MVMS-ICP

The IC signature generated in MVMS-ICP satisfies **linearity** property, which may be required in many applications of ICP (specifically in statistical VSS and MPC [2, 3]). Specifically, consider the following settings: let in $q$ different instances of MVMS-ICP, $D$ has handed over IC Signature on $q$ different set of $\ell$ secrets to $INT$, namely $S_i = (s_i^1, \ldots, s_i^\ell)$, for $i = 1, \ldots, q$. Moreover, let $D$ has used the same $\alpha_i$ as secret evaluation point for verifier $P_i$ in all the $q$ instances of MVMS-ICP (an honest $D$ can always ensure it). This condition on $\alpha_i$ is very important and we refer this as *the condition for linearity of IC signatures*. Though linearity property accounts for any form of linear function, we will demonstrate the linearity property with respect to addition operation (for simplicity). So let $S = S_1 + \ldots + S_q$, where $S = (s^1, \ldots, s^\ell)$ and $s^l = s_1^l + \ldots + s_q^l$, for $l = 1, \ldots, \ell$. Now $INT$ can compute $ICSig(D, INT, \mathcal{P}, S)$ using $ICSig(D, INT, \mathcal{P}, S_i)$ for $i = 1, \ldots, q$ and the verifiers can compute verification information corresponding to $ICSig(D, INT, \mathcal{P}, S)$, without doing any further communication. For the sake of completeness, we present a protocol in Fig. 2 showing how $INT$ and verifiers can achieve the above. Informally in the protocol we use the linearity property of polynomials. That is, if $ICSig(D, INT, \mathcal{P}, S_1) = F_1(x)$ and $ICSig(D, INT, \mathcal{P}, S_2) = F_2(x)$, then $ICSig(D, INT, \mathcal{P}, S_1 + S_2) = F_1(x) + F_2(x)$. Similarly, if $F_1(\alpha_i)$ and $F_2(\alpha_i)$ are the verification information of verifier $P_i$ corresponding to $ICSig(D, INT, \mathcal{P}, S_1)$ and $ICSig(D, INT, \mathcal{P}, S_2)$ respectively, then $F_1(\alpha_i) + F_2(\alpha_i)$ will be the verification information of verifier $P_i$ corresponding to $ICSig(D, INT, \mathcal{P}, S_1 + S_2)$.

In the protocol, it might be possible that some $ICSig(D, INT, \mathcal{P}, S_i)$ is a polynomial of degree $\ell + t$ (this implies that $D$ has not broadcasted anything during Ver of $i^{th}$ signature giving instance), while some other $ICSig(D, INT, \mathcal{P}, S_j)$ is $S_j$ (this implies that $D$ has broadcasted $S_j$ during Ver of $j^{th}$ signature giving instance). In such a case, $INT$ finds a $\ell + t$ degree polynomial $F_j(x)$, whose lower order $\ell$ coefficients are elements of $S_j$ and the remaining coefficients are some publicly known default values and assumes the polynomial to be $ICSig(D, INT, \mathcal{P}, S_j)$. Notice that such $F_j(x)$ will be known publicly, as $S_j$ is broadcasted by $D$. Accordingly, every verifier $P_i$ considers $F_j(\alpha_i)$ as his verification information corresponding to $ICSig(D, INT, \mathcal{P}, S_j)$. Once this is done then all the $q$ IC signatures will be $\ell + t$ degree polynomials and hence $INT$ can use the linearity property of the polynomials (as explained above) to compute the addition of IC signatures.

Now we show that a linearly combined IC signature that is computed from $q$ IC signatures (using protocol in Fig. 2), each having $\epsilon$ error, will have $\epsilon$ error. For this, we prove the following lemma:

**Lemma 5** *Assuming each of the $q$ individual IC signatures, $ICSig(D, INT, \mathcal{P}, S_j)$ has $\epsilon$ error, the linearly combined IC signature, $ICSig(D, INT, \mathcal{P}, S)$ will also have $\epsilon$ error.*

PROOF: We will examine each of the four properties of IC signature one by one depending on whether $D$ and/or $INT$ are honest or corrupted. When $D$ and $INT$ are honest, then it is easy to see that $ICSig(D, INT, \mathcal{P}, S)$ will abide by **ICP-Correctness1** and **ICP-Secrecy** without any error.

Now when $D$ is honest and $INT$ is corrupted, $ICSig(D, INT, \mathcal{P}, S)$ satisfies **ICP-Correctness3** with error probability $\epsilon$, which is same as the error of individual IC signatures. This is because, here the error probability depends on correctly guessing one of the honest $P_i$'s $\alpha_i$ (recall that same $\alpha_i$ is associated with $P_i$ corresponding to all the individual IC signatures).

Finally, we show that when $D$ is corrupted and $INT$ is honest, $ICSig(D, INT, \mathcal{P}, S)$ satisfies **ICP-Correctness2** with error probability $\epsilon$. The worst case that causes this error probability is:

1. To every honest verifier $P_i$, $D$ gives $v_{ji} \neq F_j(\alpha_i)$ and $r_{ji} \neq R_j(\alpha_i)$, corresponding to exactly one $j \in \{1, \ldots, q\}$;

2. For all other $j \in \{1, \ldots, q\}$, $D$ gives $v_{ji} = F_j(\alpha_i)$ and $r_{ji} = R_j(\alpha_i)$ to every honest verifier $P_i$.

In this case, from the proof of Lemma 2, $B_j(\alpha_i) \neq d_j v_{ji} + d_j r_{ji}$ will not hold for some honest $P_i$, except with probability $\epsilon$. Now notice that if $D$ delivers $v_{ji}, r_{ji}$ satisfying $v_{ji} \neq F_j(\alpha_i)$ and $r_{ji} \neq R_j(\alpha_i)$ for more $j$'s, then $D$ has to guess more $d_j$'s and hence the probability with which $D$ can guess all those $d_j$'s will decrease beyond $\epsilon$. Hence we proved that when $D$ is corrupted and $INT$ is honest, $ICSig(D, INT, \mathcal{P}, S)$ satisfies **ICP-Correctness2** with error probability $\epsilon$. Hence the lemma. $\square$

The linearity of IC signatures also captures the following case: Let in an execution of MVMS-ICP, $D$ has handed over IC Signature on a set of $\ell$ secrets to $INT$, say $b^1, \ldots, b^\ell$. That is at the end of Ver, $INT$ holds $ICSig(D, INT, \mathcal{P}, (b^1, \ldots, b^\ell))$. Also let $(a^1, \ldots, a^\ell)$ are some publicly known values. Now

Figure 2: Linearity of Protocol MVMS-ICP Over Addition Operation.

---

**Assumption:**

1. $D$ has sent $ICSig(D, INT, \mathcal{P}, S_j)$ having $\epsilon$ error to $INT$, for $j = 1, \ldots, q$, where $S_j = (s_j^1, \ldots, s_j^\ell)$. Let $D$ has used the same $\alpha_i$ as secret evaluation point for verifier $P_i$ in all the $q$ instances for giving IC signatures. Moreover, let $INT$ has used random value $d_j$ in **Round 1** of Ver for $j^{th}$ signature giving instance of MVMS-ICP.

2. $INT$ has received $ICSig(D, INT, \mathcal{P}, S_j)$ having $\epsilon$ error from $D$.

3. For every $j \in \{1, \ldots, q\}$, such that $ICSig(D, INT, \mathcal{P}, S_j)$ is a polynomial of degree $\ell + t$, let $ICSig(D, INT, \mathcal{P}, S_j) = F_j(x)$, i.e $D$ had used $F_j(x)$ to hide $S_j$. Moreover let $P_i$ has the verification information $v_{ji}$, which is supposed to be same as $F_j(\alpha_i)$.

**Local Computation to Compute Addition of IC Signatures:**

1. For all $j \in \{1, \ldots, q\}$, such that $ICSig(D, INT, \mathcal{P}, S_j) = S_j$, $INT$ assumes a degree $\ell + t$ polynomial $F_j(x)$ whose lower order $\ell$ coefficients are the elements of $S_j$ and the remaining coefficients are some publicly known default values. Notice that such $F_j(x)$ polynomials will be known publicly. For every such $F_j(x)$, verifier $P_i$ computes his verification information as $v_{ji} = F_j(\alpha_i)$.

2. Now to compute $ICSig(D, INT, \mathcal{P}, S)$, $INT$ sets $F(x) = \sum_{j=1}^{q} F_j(x)$ and assigns $ICSig(D, INT, \mathcal{P}, S) = F(x)$.

3. Every verifier $P_i$ computes his verification information corresponding to $ICSig(D, INT, \mathcal{P}, S)$ in the following way: $v_i = \sum_{j=1}^{q} v_{ji}$.

**Revelation of Linear IC Signature:**

1. $INT$ broadcasts $ICSig(D, INT, \mathcal{P}, S)$ (i.e $F(x)$).

2. Verifier $P_i$ broadcasts `Accept` if one of the following conditions holds.

   (a) **C1:** $v_i = F(\alpha_i)$; OR
   
   (b) **C2:** For some $j \in \{1, \ldots, q\}$, $B_j(\alpha_i) \neq d_j v_{ji} + r_{ji}$ ($B_j(x)$ was broadcasted by $INT$ during **Round 1** of Ver of $j^{th}$ signature giving instance) and $D$ has not broadcasted $S_j$ in **Round 2** of Ver of $j^{th}$ signature giving instance.

   Otherwise, $P_i$ broadcasts `Reject`.

**Local Computation (By Every Verifier)**: If at least $(t+1)$ verifiers have broadcasted `Accept` then accept $ICSig(D, INT, \mathcal{P}, S)$ and hence $S$. Else reject $ICSig(D, INT, \mathcal{P}, S)$.

---

$INT$ can compute $ICSig(D, INT, \mathcal{P}, (b^1 - a^1, \ldots, b^\ell - a^\ell))$ and similarly verifiers can update their verification information accordingly, by doing local computation. Later in Reveal, $INT$ can reveal $ICSig(D, INT, \mathcal{P}, (b^1 - a^1, \ldots, b^\ell - a^\ell))$ to the verifiers. Moreover, the above idea can be extended for any number of IC signatures and any number of sets containing publicly known values.

**Note 1** *We would like to alert that linearity of IC signatures holds only when all the IC signatures are generated by same party, say $P$ (who acts as a dealer). Moreover, $P$ should abide by the condition for the linearity of IC signatures. Linearity does not hold on the IC signatures that are generated by different parties, as they will not satisfy condition for the linearity of IC signatures (because different parties may choose different $\alpha_i$ for verifier $P_i$ in their signature giving instance).*

# 6 Conclusion and Open Problems

In this paper, we have extended the basic bare-bone definition of ICP, introduced by Rabin et al. [3] and subsequently followed by [1, 2], to capture multiple verifiers and multiple secrets concurrently. Then we have presented a novel ICP (matching with our definition) that turns out to be the best ICP in the literature as per the round and communication complexity. We then showed that our ICP satisfies the linearity property. We now conclude this paper with the following interesting open questions: Can we improve the round and communication complexity of MVMS-ICP when $n = 2t + 1$?

This leads to a more general question: What is the round and communication complexity lower bound for ICP with $n = 2t + 1$ verifiers? ICP can be studied in multi verifier and multi secret settings in asynchronous network where we may investigate the issues like communication efficiency etc. An initiative in this direction has been taken in [4].

**Acknowledgement**: We would sincerely like to thank Tal Rabin for several fruitful discussions.

# References

[1] R. Canetti and T. Rabin. Fast Asynchronous Byzantine Agreement with Optimal Resilience. In *STOC*, pages 42–51. ACM Press, 1993.

[2] R. Cramer, I. Damgård, S. Dziembowski, M. Hirt, and T. Rabin. Efficient Multiparty Computations Secure Against an Adaptive Adversary. In *EUROCRYPT*, LNCS 1592, pages 311–326. Springer, 1999.

[3] T. Rabin and M. Ben-Or. Verifiable Secret Sharing and Multiparty Protocols with Honest Majority. In *STOC*, pages 73–85. ACM Press, 1989.

[4] A. Patra, A. Choudhary and C. Pandu Rangan. Efficient Asynchronous Byzantine Agreement with Optimal Resilience. In *PODC*, pages 92-101. ACM Press, 2009.